\documentclass[12pt]{article}

\usepackage{amsfonts}

\usepackage{amssymb}
\usepackage{latexsym}
\usepackage{graphicx}
\usepackage[english]{babel}
\usepackage[font={small}]{caption}

\usepackage{amsfonts}
\usepackage{latexsym}
\usepackage{graphicx}
\usepackage[english]{babel}
\usepackage{tikz}

\begin{document}
\input epsf

\def\p{\partial}
\def\h{{1\over 2}}
\def\be{\begin{equation}}
\def\bea{\begin{eqnarray}}
\def\ee{\end{equation}}
\def\eea{\end{eqnarray}}
\def\d{\partial}
\def\la{\lambda}
\def\eps{\epsilon}
\def\bb{\bigskip}
\def\mm{\medskip}
\newcommand{\dm}{\begin{displaymath}}
\newcommand{\edm}{\end{displaymath}}
\renewcommand{\b}{\tilde{B}}
\newcommand{\gm}{\Gamma}
\newcommand{\ac}[2]{\ensuremath{\{ #1, #2 \}}}
\renewcommand{\ell}{l}
\newcommand{\z}{\ell}
\newcommand{\newsection}[1]{\section{#1} \setcounter{equation}{0}}
\def\bb{$\bullet$}
\def\Qbar{{\bar Q}_1}
\def\QPbar{{\bar Q}_p}

\def\q{\quad}

\def\bn{B_\circ}

\let\a=\alpha \let\b=\beta \let\g=\gamma \let\d=\delta \let\e=\epsilon
\let\c=\chi \let\th=\theta  \let\k=\kappa
\let\l=\lambda \let\m=\mu \let\n=\nu \let\x=\xi \let\r=\rho
\let\s=\sigma \let\t=\tau
\let\vp=\varphi \let\vep=\varepsilon
\let\w=\omega      \let\G=\Gamma \let\D=\Delta \let\Th=\Theta
                     \let\P=\Pi \let\S=\Sigma

\def\h{{1\over 2}}
\def\t{\tilde}
\def\r{\rightarrow}
\def\nn{\nonumber\\}
\let\bm=\bibitem
\def\Kt{{\tilde K}}
\def\b{\bigskip}

\let\p=\partial

\begin{flushright}
\end{flushright}
\vspace{20mm}
\begin{center}
{\LARGE Spacetime has a `thickness'  \footnote{Essay awarded second prize in the Gravity Research Foundation 2017 Awards for Essays on Gravitation.}}
\\
\vspace{18mm}
 Samir D. Mathur

\vskip .1 in

 Department of Physics\\The Ohio State University\\ Columbus,
OH 43210, USA\\mathur.16@osu.edu\\
\vspace{4mm}
 March 31, 2017
\end{center}
\vspace{10mm}
\thispagestyle{empty}
\begin{abstract}

Suppose we assume that (a) information about a black hole is encoded in its Hawking radiation and (b) causality is not violated to leading order in gently curved spacetime.  Then we argue that spacetime cannot just be described as a manifold with a shape; it must be given an additional attribute which we call `thickness'. This thickness characterizes the spread of the  quantum gravity wavefunctional  in superspace -- the space of all 3-geometries. Low energy particles travel on spacetime without noticing the thickness parameter, so they just see an effective manifold.  Objects with energy large enough to create a horizon do notice the finite thickness; this modifies the semiclassical evolution in such a way that we  avoid  horizon formation and the consequent violation of causality.

\end{abstract}
\vskip 1.0 true in

\newpage
\setcounter{page}{1}

Start with a black hole of mass $M$
\be
ds^2=-(1-{2GM\over r}) dt^2 + {dr^2\over 1-{2GM\over r}} + r^2(d\theta^2+\sin^2\theta d\phi^2)
\label{one}
\ee
Let a spherical shell carrying energy $\Delta M$ be incident on this hole. Consider two cases:

\b

(i) The limit $\Delta M\r 0$ where the shell travels as a test particle on the background (\ref{one}). In this case there is no contradiction if we assume that  the shell reaches the horizon at $r=r_h=2GM$  without encountering any significant effects of quantum gravity.

\b

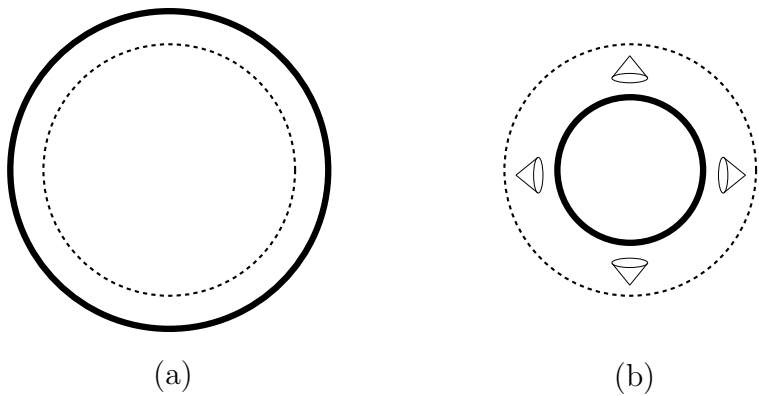
\begin{figure}
\hskip 1 in
\begin{tikzpicture}[y=0.80pt, x=.80pt, yscale=-.156000000, xscale=.156000000, inner sep=0pt, outer sep=0pt]
  \begin{scope}[shift={(0,-48.57141)}]
    \path[draw=black,dash pattern=on 1.60pt off 1.60pt,line join=miter,line
      cap=butt,miter limit=4.00,even odd rule,line width=0.800pt]
      (368.5714,580.9336) circle (10.7244cm);
    \path[draw=black,line join=miter,line cap=butt,miter limit=4.00,even odd
      rule,line width=2.400pt] (368.5714,580.9336) circle (6.2089cm);
    \path[draw=black,line join=miter,line cap=butt,miter limit=4.00,even odd
      rule,line width=0.283pt] (367.1428,302.1688) ellipse (1.5232cm and 0.3993cm);
    \path[draw=black,line join=miter,line cap=butt,miter limit=4.00,even odd
      rule,line width=0.283pt] (313.9304,300.1467) -- (367.5256,234.0937) --
      (367.5256,234.0937) -- (367.5256,234.0937) -- (367.5256,234.0937);
    \path[draw=black,line join=miter,line cap=butt,miter limit=4.00,even odd
      rule,line width=0.283pt] (421.0006,300.8207) -- (367.4054,234.7677) --
      (367.4054,234.7677) -- (367.4054,234.7677) -- (367.4054,234.7677);
    \path[xscale=1.000,yscale=-1.000,draw=black,line join=miter,line cap=butt,miter
      limit=4.00,even odd rule,line width=0.283pt] (367.1428,-862.5269) ellipse
      (1.5232cm and 0.3993cm);
    \path[draw=black,line join=miter,line cap=butt,miter limit=4.00,even odd
      rule,line width=0.283pt] (313.9304,864.5489) -- (367.5256,930.6019) --
      (367.5256,930.6019) -- (367.5256,930.6019) -- (367.5256,930.6019);
    \path[draw=black,line join=miter,line cap=butt,miter limit=4.00,even odd
      rule,line width=0.283pt] (421.0006,863.8749) -- (367.4054,929.9279) --
      (367.4054,929.9279) -- (367.4054,929.9279) -- (367.4054,929.9279);
    \path[cm={{0.0,-1.0,1.0,0.0,(0.0,0.0)}},draw=black,line join=miter,line
      cap=butt,miter limit=4.00,even odd rule,line width=0.283pt]
      (-596.6335,89.8208) ellipse (1.5232cm and 0.3993cm);
    \path[draw=black,line join=miter,line cap=butt,miter limit=4.00,even odd
      rule,line width=0.283pt] (87.7988,649.8459) -- (21.7458,596.2507) --
      (21.7458,596.2507) -- (21.7458,596.2507) -- (21.7458,596.2507);
    \path[draw=black,line join=miter,line cap=butt,miter limit=4.00,even odd
      rule,line width=0.283pt] (88.4728,542.7757) -- (22.4198,596.3709) --
      (22.4198,596.3709) -- (22.4198,596.3709) -- (22.4198,596.3709);
    \path[cm={{0.0,-1.0,-1.0,0.0,(0.0,0.0)}},draw=black,line join=miter,line
      cap=butt,miter limit=4.00,even odd rule,line width=0.283pt]
      (-596.6335,-650.1790) ellipse (1.5232cm and 0.3993cm);
    \path[draw=black,line join=miter,line cap=butt,miter limit=4.00,even odd
      rule,line width=0.283pt] (652.2010,649.8459) -- (718.2540,596.2507) --
      (718.2540,596.2507) -- (718.2540,596.2507) -- (718.2540,596.2507);
    \path[draw=black,line join=miter,line cap=butt,miter limit=4.00,even odd
      rule,line width=0.283pt] (651.5270,542.7757) -- (717.5800,596.3709) --
      (717.5800,596.3709) -- (717.5800,596.3709) -- (717.5800,596.3709);
  \end{scope}
  \begin{scope}[shift={(2.85714,-40.0)}]
    \path[draw=black,dash pattern=on 1.60pt off 1.60pt,line join=miter,line
      cap=butt,miter limit=4.00,even odd rule,line width=0.800pt]
      (-1031.4286,572.3622) circle (10.7244cm);
    \path[draw=black,line join=miter,line cap=butt,miter limit=4.00,even odd
      rule,line width=2.400pt] (-1031.4286,572.3622) circle (13.5467cm);
  \end{scope}
  \path[fill=black,line join=miter,line cap=butt,line width=0.800pt]
    (-1077.4313,1204.3622) node[above right] (text4470) {(a)};
  \path[fill=black,line join=miter,line cap=butt,line width=0.800pt]
    (319.0280,1212.9629) node[above right] (text4474) {(b)};

\end{tikzpicture}

\caption{(a) A shell of mass $M$ is collapsing towards its horizon. (b) If the shell passes through its horizon, then the information it carries is trapped inside the horizon due to the structure of light cones.} \label{fig1}

\end{figure}

(ii) The case where $\Delta M$ is not infinitesimal. In this case semiclassical evolution implies that the shell will create a new horizon at
\be
\t r_h = 2G(M+\Delta M)
\label{two}
\ee
and continue its passage to smaller $r$. Light cones in the region $r<\t r_h$ now point inwards, as depicted in fig.\ref{fig1}. Suppose we assume that causality holds in our theory; i.e., information cannot travel outside the light cones. Then how will the information in the shell ever return to infinity?

\b

One might think that Hawking radiation \cite{hawking} is a quantum effect, and can somehow encode the information of the shell in delicate correlations among the emitted quanta. But it was shown in \cite{cern} that this is not possible, if we assume that the leading order evolution of low energy modes around the horizon is described by semiclassical physics. Suppose  the unknown quantum gravity effects alter the semiclassical evolution of Hawking modes by a fraction $\epsilon\ll 1$. Then using the strong subadditivity of quantum entanglement entropy it can be shown that the fraction $f$ of information that can be encoded in Hawking radiation is bounded at all times as
\be
f\equiv {\delta S_{ent}\over S_{ent}} < 2\epsilon
\ee
where $S_{ent}$ is the entanglement entropy of  the radiated quanta with the remaining hole.

If small corrections to semiclassical evolution at the horizon will not solve the problem, then we need $O(1)$ corrections. And indeed, as we note below, in string theory we find that black hole microstates are `fuzzballs', which have a physical size $r_f\approx 2GM+l_p$ and thus no horizon at all \cite{fuzzballs}. But this leaves open a crucial question: {\it what tells the incoming shell that it should transition to a fuzzball at the location $\t r_h$?}

\b

If spacetime were just a manifold, then we have only two choices: 

\b

(a) Assume that the  metric (\ref{one}) does {\it not} describe evolution at $r=\t r_h$. Then we can say that this altered spacetime has some feature  -- for example a curvature singularity at $\t r_h$ -- that will trigger fuzzball formation when the shell approaches  $r=\t r_h$. But by considering arbitrarily large $\Delta M$, we can reach this conclusion for arbitrarily large $\t r_h$. This does not make sense, since the effects of the black hole should not extend to arbitrarily large distances from the hole. 

\b

(b) Assume that the metric (\ref{one}) does describe the spacetime at $r=\t r_h$. Since the curvature at $\t r_h$  is much lower than planck scale,  we can go to coordinates in which the vicinity of any point $\t r, \theta, \phi$ is a patch of Minkowski space. Then the equivalence principle tells us that the shell will uneventfully pass into the region $r<\t r_h$. In this case we  will face the causality problem mentioned above.

\b

\begin{figure}\label{bau}
 \includegraphics[scale=.16] {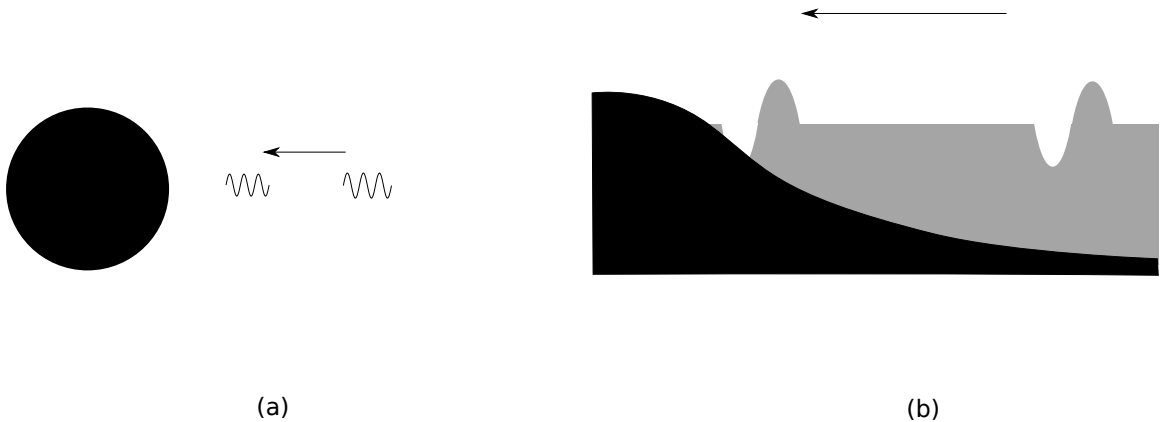}
\caption{\label{fig2q} (a) The fuzzball (black region) and a quantum coming in from infinity. (b) A model describing the quantum nature of spacetime. The black region is land (interior of the fuzzball), while the grey region is water (the region outside the fuzzball).  A wave of  travels freely when its amplitude is much less than the depth of the water, but will suffer nontrivial deformation when its amplitude becomes comparable to the depth of the water. } 
\end{figure}

The only way out is to have completely  different evolutions in the cases  (i) and (ii) described above; i.e., have spacetime behave differently for the propagation of shells with $\Delta M$ small and $\Delta M$ large.  Thus spacetime must have an additional property  that distinguishes these possibilities; we will call this property `thickness' or `depth'. The essential idea is pictured in fig.\ref{fig2q}(b). Think of spacetime as a lake. The dark area on the left is land; this represents the fuzzball which describes a microstate of the black hole in the region  $r<2GM +l_p$ (fig.\ref{fig2q}(a)). On the right is water, and waves on this water describe the propagation of a scalar field $\phi$ in the region $r>2GM+l_p$.  For our two cases (i) and (ii) above we have the following:

\b

(i') A low energy excitation of $\phi$  (i.e. $\Delta M \r 0$) is described by a wave with small amplitude. Such  a wave does not notice the finite depth of the lake and travels with the semiclassical evolution $\square \phi=0$ upto $r\approx r_h$. 

\b

(ii') A  high energy excitation of $\phi$  is described by a wave with large amplitude. As this wave comes in from large $r$, it touches the bottom of the lake at some point $\t r_h$; the value of $\t r_h$ is related to the energy $\Delta M$ of the incoming excitation through (\ref{two}). The evolution of the scalar is therefore modified at $r\approx \t r_h$, departing from the semiclassical expectation. As we will discuss below, this modification will yield a fuzzball rather than a horizon. In this way we avoid the conflict with causality depicted in fig.\ref{fig1}. 

The notion of  a `finite depth sea' appeared earlier in the exact solution of a simple string theory, called the $c=1$ model \cite{dasjevicki}. This model has a scalar field $\phi$ which is described as `waves on a fermi sea'. The gravity theory described by this model does not have a black hole.  But  it is interesting that  the energy threshold corresponding to the  wave touching the bottom of the sea is of the {\it order} required to make a black hole \cite{polnat}. In full string theory, the study of simple black holes suggested that spacetime should be thought of as  a rubber sheet which becomes thinner as it is stretched; when the sheet becomes  thin, it is no longer able to act as a smooth manifold for particles with wavelength  smaller than the spacing between the atoms making the sheet \cite{universe}.

\begin{figure}
 \includegraphics[scale=.14] {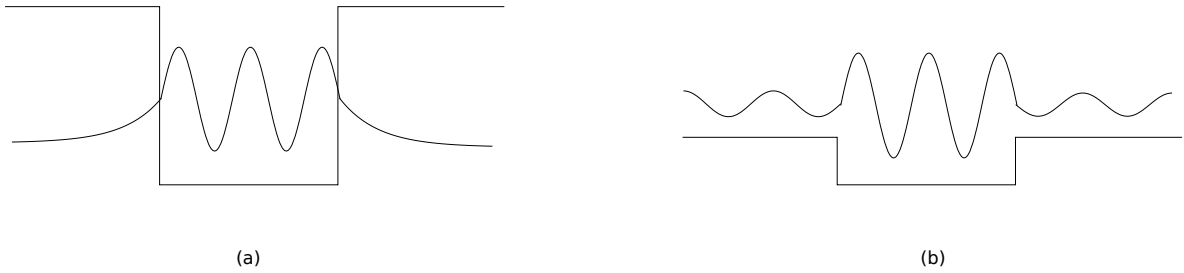}
\caption{\label{fig3q} A schematic depiction of superspace. (a) The wavefunctional is oscillatory in the region of low potential; this oscillatory part describes  semiclassical gravity. The wavefunction is damped in the region of high potential, which corresponds to fuzzball configurations with mass more than the available energy. (b) An infalling shell lowers the barrier, creating fuzzballs as real rather than virtual excitations.  }
\end{figure}

But what is the origin of this new parameter -- `thickness' or `depth' -- that spacetime must possess? In string theory there are $Exp[S_{bek}]$ horizon sized microstates called fuzzballs which account for the entropy of the hole. Even in empty spacetime, there are {\it virtual fluctuations} corresponding to the deformation of space into one of these fuzzball configurations. If we start with a black hole (i.e. a fuzzball state) of mass $M$, then there will be similar fluctuations to fuzzballs of mass $M'>M$. To describe these fluctuations, we must study the wavefunctional  in superspace -- the space of all configurations of our gravity theory. The virtual fluctuations are represented by the part of the wavefunction that is `under the barrier'; i.e., the part which tunnels to regions of superspace that cannot be on-shell for the given energy of the state. This situation is depicted in fig.\ref{fig3q}. Superspace has a very large number of directions, but we have depicted the allowed and `under barrier' regions using a 1-d potential well. 

We conjecture that {\it the `thickness' of spacetime describes the height of the potential barrier in the forbidden region of superspace}. For our cases (i) and (ii) we have the following situation:

\b

(i'') Consider a low energy shell $\Delta M \r 0$ approaching the black hole. For such low energy excitations the forbidden region -- the part under the barrier -- is largely irrelevant. The dynamics is given to leading order by the evolution of the wavefunction in the `allowed' region; i.e., the region where the wavefunctional is oscillatory rather than damped (fig.\ref{fig3q}(a)). 

\b

(ii'') Now consider the infall of a heavy shell with energy $\Delta M$. When this shell is at a radius $r$, it creates a negative potential 
\be
\Phi\approx -{GM\over r} 
\ee
in the region outside the shell. This lowers the energy of virtual excitations, in particular in the region just outside the shell.  This is depicted in fig.\ref{fig3q}(b). When $r$ reaches a critical value $\t r_h$, the potential barrier is lowered enough for fuzzball configurations with mass $M+\Delta M$ to become `allowed'; i.e, the wavefunctional at this point in superspace becomes oscillatory. This corresponds to the shell transitioning into fuzzballs, and we then evade the information paradox and the associated causality problem \cite{flaw,causality}. 

\b

It may appear strange that quantum gravity effects can alter the semiclassical evolution of the shell at a macroscopic radius $r=\t r_h$. Indeed the amplitude for the shell to transition to any one fuzzball configuration is small
\be
{\cal P} \sim e^{-2S_{cl}}\ll 1
\ee
where $S_{cl}$ is the classical action for the tunneling process.
 But this smallness is offset by the large degeneracy of fuzzball states
 ${\cal N} \sim Exp[{S_{bek}}]$
\be
{\cal P} {\cal N} \sim e^{-2S_{cl}}e^{S_{bek}}\sim 1
\ee
to yield an order unity correction to semiclassical dynamics \cite{tunnel}. Thus the large value of $S_{bek}$ -- a characteristic feature of black holes -- is responsible for the quantum effects we are postulating.

To summarize, all problems with black holes arise because the metric (\ref{one}) suggests that nothing special happens when a shell of mass $\Delta M$ passes through its horizon radius (\ref{two}).  We have argued that this problem is resolved by giving spacetime a new attribute -- `thickness'. Traditional semiclassical physics incorporating the equivalence principle holds when the energies are low enough that the wave in fig.\ref{fig2q}(b) does not feel the bottom of the lake. The situation changes at the black hole threshold where we do hit the bottom of the lake; the shell then transitions to fuzzballs.  This thickness arises naturally as the spread of the wavefunctional in the directions of superspace that are `under the barrier'; these directions are very large in number and correspond to fuzzballs with mass exceeding the available energy. The infalling shell lowers the barrier so that the virtual fuzzballs become real; this prevents horizon formation and resolves the information paradox.  

\b

\section*{Acknowledgements}

I am grateful to Borun Chowdhury,  Sumit Das, A. Jevicki, Oleg Lunin, Emil Martinec and  David Turton for helpful discussions. This work is supported in part by a grant from the FQXi foundation.

\end{document}